\documentclass[a4paper, 12pt]{article}

\usepackage{amsmath}
\usepackage{graphicx}
\usepackage{psfrag}

\begin{document}
\renewcommand{\theequation}{\thesection .\arabic{equation}}

\newcommand{\sign}{\operatorname{sign}}
\newcommand{\Ci}{\operatorname{Ci}}
\newcommand{\tr}{\operatorname{tr}}

\newcommand{\beq}{\begin{equation}}
\newcommand{\eeq}{\end{equation}}
\newcommand{\beqn}{\begin{eqnarray}}
\newcommand{\eeqn}{\end{eqnarray}}

\newcommand{\slp}{\raise.15ex\hbox{$/$}\kern-.57em\hbox{$ \partial $}}
\newcommand{\lnA}{\raise.15ex\hbox{$/$}\kern-.57em\hbox{$A$}}
\newcommand{\unmedio}{{\scriptstyle\frac{1}{2}}}
\newcommand{\uncuarto}{{\scriptstyle\frac{1}{4}}}

\newcommand{\trial}{_{\text{trial}}}
\newcommand{\true}{_{\text{true}}}
\newcommand{\const}{\text{const}}

\newcommand{\intp}{\int\frac{d^2p}{(2\pi)^2}\,}
\newcommand{\intx}{\int d^2x\,}
\newcommand{\inty}{\int d^2y\,}
\newcommand{\intxy}{\int d^2x\,d^2y\,}

\newcommand{\bP}{\bar{\Psi}}
\newcommand{\bc}{\bar{\chi}}
\newcommand{\hs}{\hspace*{0.6cm}}

\newcommand{\bra}{\left\langle}
\newcommand{\ket}{\right\rangle}
\newcommand{\bracket}{\left\langle\,\right\rangle}

\newcommand{\D}{\mbox{$\mathcal{D}$}}
\newcommand{\N}{\mbox{$\mathcal{N}$}}
\newcommand{\Lag}{\mbox{$\mathcal{L}$}}
\newcommand{\V}{\mbox{$\mathcal{V}$}}
\newcommand{\Z}{\mbox{$\mathcal{Z}$}}
\renewcommand{\textfraction}{0.08}

\begin{titlepage}

%\begin{flushright}
%La Plata-Th 00/01\\
%\end{flushright}

\vspace{2cm}

\begin{center}

{\Large {\bf Neutrino Quasinormal Modes of a Kerr-Newman-de Sitter Black Hole}}

\vspace{1.3cm}

Jia-Feng Chang$^{a,b,c}$ and You-Gen Shen$^{a,c,d}$\footnote{e-mail: ygshen
@center.shao.ac.cn}

\vspace{.8cm}

$^a$ {\it Shanghai Astronomical Observatory, Chinese Academy of
Sciences, Shanghai 200030, China.}

\smallskip

$^b$ {\it Graduate School of Chinese Academy of Sciences,Beijing
100039,China.}

\smallskip

$^c$ {\it National Astronomical Observatories, Chinese Academy of
Sciences, Beijing 100012, China.}

\smallskip
$^d$ {\it Institute of Theoretical physics, Chinese Academy of
Sciences, Beijing 100080, China}

\vspace{1cm}

\begin{abstract}
Using the P\"{o}shl-Teller approximation, we evaluate the neutrino quasinormal modes
(QNMs) of a Kerr-Newman-de Sitter black hole. The result shows that for a
Kerr-Newman-de Sitter black hole, massless neutrino perturbation of large $\Lambda$,
positive $m$ and small value of $n$ will decay slowly.
 \end{abstract}
\end{center}

\vspace{1 cm}

\noindent{\it PACS number(s):} 04.70.-s, 04.50.+h, 11.15.-q, 11.25.Hf

\end{titlepage}

\newpage
\section{Introduction}
\hs It is well-known that there are three stages during the evolution of the field
perturbation in the black hole background: the initial outburst from the source of
perturbation, the quasinormal oscillations and the asymptotic tails. The frequencies
and damping time of the quasinormal oscillations called "quasinormal modes"(QNMs) are
determined only by the black hole's parameters and independent of the initial
perturbations. A great deal of efforts have been devoted to the black hole's QNMs for
the possibility of direct identification of black hole existence through gravitational
wave detectors in the near future \cite{qnmr1,qnmr2}. The study of black hole's QNMs
has a long history. Most of the studies immersed in an asymptotically flat space time.
The discovery of the $AdS/CFT$ \cite{birmingham,cardoso2} correspondence and the
expanding universe motivated the investigation of QNMs in de Sitter
\cite{dsitter1,dsitter2} and anti-de Sitter
\cite{adsitter1,adsitter2,adsitter21,adsitter3}
space time in the past several years.\\
\hspace*{7.5mm}Most methods in evaluating the QNMs are numerical in nature. Recently,
using the third-order WKB approximation, Cho evaluated the Dirac field QNMs of a
Schwarzschild black hole \cite{cho}. A powerful WKB scheme was devised by Schutz and
Will \cite{Wkb1}, and was extended to higher orders in \cite{Wkb2}. Konoplya
\cite{konoplya} extended the WKB approximation to sixth-order and calculated the QNMs
of a D-dimensional Schwarzshild black hole. Zhidenko \cite{zhidenko} calculated
low-laying QNMs of a Schwarzschild-de Sitter black hole by using sixth-order WKB
approximation and the approximation by P\"{o}shl-Teller potential. Cardoso
\cite{cardoso} calculated QNMs of the near extremal Schwarzschild-de Sitter black hole
by using P\"{o}shl-Teller approximation, which was proved to be exactly in the near
extreme regime \cite{cardoso1}. Yoshida \cite{dsitter3} numerically analyzed QNMs in
nearly extremal Schwarzschild-de Sitter spacetimes. The Kerr black hole is a more
general case. It is also important to note that the most important QNMs are the lowest
ones which have smaller imaginary on the astrophysical aspect and the most important
spacetimes are the asymptotically flat and now perhaps the asymptotically de Sitter
which supported by the recent observation data. So we discuss the QNMs of a
Kerr-Newman-de Sitter black hole in this paper. Leaver \cite{leaver} developed a hybrid
analytic-numerical method to calculate the QNMs of black holes and applied to the Kerr
black hole. Seidel and Iyer \cite{Wkb3} computed the low-laying QNMs of Kerr black
holes for both scalar and gravitational perturbations by using third-order WKB
approximation.
 Berti et.al dealt with highly damped QNMs of Kerr black holes in \cite{kerr1,kerr2}.\\
\hspace*{7.5mm}In this paper, we evaluate the QNMs of Kerr-Newman-de Sitter black hole
for neutrino perturbation. In Section \textbf{2} we consider the massless Dirac
equations for massless neutrino in the Kerr-Newman-de Sitter black hole and reduced it
into a set of Schr\"{o}dinger-like equations with a particular effective potential. We
analyse the properties of the particular potential in section \textbf{3} and use the
P\"{o}shl-Teller potential approximation to evaluate the QNMs of massless neutrino in
Section \textbf{4}. Conclusions and discussions are presented in Section \textbf{5}.
Throughout this paper we use units in which $G=c=M=1$.

\section{MASSLESS DIRAC FIELD EQUATION IN THE KERR-NEWMAN-DE SITTER BLACK HOLE}
\setcounter{equation}{0}

\hs Generally speaking, neutrino is a kind of uncharged Dirac particles without rest
mass or with tiny mass. In curved spacetime, the spinor representations of massless
Dirac equations are \cite{chandrasekhar}
\begin{eqnarray}\label{dirac1}
\nabla_{A\dot{B}}P^{A}=0,\\
\label{dirac2}\nabla_{A\dot{B}}Q^{A}=0,
\end{eqnarray}
where $P^{A}$ and $Q^{A}$ are two two-component spinors, the operator
$\nabla_{A\dot{B}}$ denotes the spinor covariant differentiation.
$\nabla_{A\dot{B}}=\sigma^{\mu}_{A\dot{B}}\nabla_{\mu}$, and $\sigma^{u}_{A\dot{B}}$
are $2\times2$ Hermitian matrices which satisfy
$g_{\mu\nu}\sigma^{\mu}_{A\dot{B}}\sigma^{\nu}_{C\dot{D}}=\epsilon_{AC}\epsilon_{\dot{B}\dot{D}}$,
where $\epsilon_{AC}$ and $\epsilon_{\dot{B}\dot{D}}$ are antisymmetric Levi-Civita
symbols, the operator $\nabla_{\mu}$ is covariant differentiation.\\
\hspace*{7.5mm}The metric of the Kerr-Newman-de Sitter black hole in the Boyer-Lindqust
coordinate system is
\begin{eqnarray}\label{metric}
ds^2&=&\frac{1}{\varrho^{2}\Sigma^2}\left(\Delta_{r}-\Delta_{\theta}a^{2}\sin^{2}\theta\right)dt^{2}
-\frac{\varrho^{2}}{\Delta_{r}}dr^{2}-\frac{\varrho^{2}}{\Delta_{\theta}}d\theta^{2}\nonumber\\
&&-\frac{1}{\varrho^{2}\Sigma^{2}}\left[\Delta_{\theta}\left(r^{2}+a^{2}\right)^{2}-
\Delta_{r}a^{2}\sin^{2}\theta\right]\sin^{2}\theta d\varphi^{2}\nonumber\\
&&+\frac{2a}{\varrho^{2}\Sigma^{2}}\left[\Delta_{\theta}{\left(r^{2}+a^{2}\right)}-\Delta_{r}\right]\sin^{2}\theta
dtd\varphi,
\end{eqnarray}
where
\begin{eqnarray}\label{metric_exp}
\bar{\varrho}&=&r+ia\cos\theta,\;\;\;\;\;\;\;\;\;\;\;\;\;\;\;\;\;\;\varrho^{2}=\bar{\varrho}\bar{\varrho}^{\ast},\nonumber\\
\Delta_{r}&=&\left(r^{2}+a^{2}\right)\left(1+\frac{\Lambda}{3}r^{2}\right)-2Mr+Q^{2},\\
\Delta_{\theta}&=&1+\frac{1}{3}\Lambda
a^2\cos^{2}\theta,\;\;\;\;\;\;\Sigma=1+\frac{1}{3}\Lambda
a^2.\nonumber\\
\end{eqnarray}
Here, $a$ and $Q$ are the angular momentum per unit mass and electric charge of the
black hole, $M$ is the black hole
mass and $\Lambda$ is the positive cosmological constant.\\
\hspace*{7.5mm}The contravariant component of metric tensor is
\begin{equation}
g^{\mu\nu}=\begin{pmatrix}\frac{\Sigma^{2}}{\varrho^{2}}\left[\frac{\left(r^{2}+a^{2}\right)^{2}}
{\Delta_{r}}-\frac{a^{2}\sin^{2}\theta}{\Delta_{\theta}}\right]&0&0&\frac{a^{2}\Sigma^{2}}{\varrho^{2}}
\left(\frac{r^{2}+a^{2}}{\Delta_{r}}-\frac{1}{\Delta_{\theta}}\right)\\0&-\frac{\Delta_{r}}{\varrho^{2}}
&0&0\\0&0&-\frac{\Delta_{\theta}}{\varrho^{2}}&0\\\frac{a^{2}\Sigma^{2}}{\varrho^{2}}
\left(\frac{r^{2}+a^{2}}{\Delta_{r}}-\frac{1}{\Delta_{\theta}}\right)&0&0&-\frac{\Sigma^{2}}{\varrho^{2}\sin^{2}
\theta}\left(\frac{1}{\Delta_{\theta}}-\frac{a^{2}\sin^{2}\theta}{\Delta_{r}}\right)
\end{pmatrix}.
\end{equation}
Choose the null tetrad as follows:
\begin{eqnarray}
l^{\mu}&=&\left[\frac{\left(r^{2}+a^{2}\right)\Sigma}{\Delta_{r}},1,0,\frac{a\Sigma}{\Delta_{r}}\right],\nonumber\\
n^{\mu}&=&\frac{1}{2\varrho^{2}}\left[\Sigma\left(r^{2}+a^{2}\right),-\Delta_{r},0,a\Sigma\right],\nonumber\\
m^{\mu}&=&\frac{1}{\sqrt{2\Delta_{\theta}}\bar{\varrho}}\left[ia\Sigma\sin\theta,0,\Delta_{\theta}
\frac{i\Sigma}{\sin\theta}\right],\nonumber\\
\bar{m}^{\mu}&=&\frac{1}{\sqrt{2\Delta_{\theta}}\bar{\varrho}^{\ast}}\left[-ia\Sigma\sin\theta,0,\Delta_{\theta},
\frac{-i\Sigma}{\sin\theta}\right].
\end{eqnarray}\\
\hspace*{7.5mm}The above null tetrad consists of null vector, i.e.
\begin{equation}
l_{\mu}l^{\mu}=n_{\mu}n^{\mu}=m_{\mu}m^{\mu}=0.
\end{equation}
The null vector satisfies the following pseudo-orthogonality relations
\begin{equation}
l_{\mu}n^{\mu}=-m_{\mu}\bar{m}^{\mu}=1,\;\;\;\;\;
l_{\mu}m^{\mu}=l_{\mu}\bar{m}^{\mu}=n_{\mu}m^{\mu}=n_{\mu}\bar{m}^{\mu}=0.
\end{equation}
They also satisfy metric conditions
\begin{equation}
g_{\mu\nu}=l_{\mu}n_{\nu}+n_{\mu}l_{\nu}-m_{\mu}\bar{m}_{\nu}-\bar{m}_{\mu}m_{\nu}.
\end{equation}
\hspace*{7.5mm}Set spinor basis $\zeta^{A}_{a}=\sigma^{A}_{a}$, in which $A$ is the
spinor component index, $a$ is the spinor basis index, both indices are 0 or 1.\\
\hspace*{7.5mm}The covariant differentiation $\nabla_{A\dot{B}}\xi^{A}$ for an
arbitrary spinor $\xi^{A}$ can be repreaented as the component along the spinor basis
$\zeta^{A}_{a}$, i.e.
\begin{equation}
\zeta^{A}_{a}\zeta^{B}_{b}\zeta^{c}_{C}\nabla_{A\dot{B}}\xi^{C}=\nabla_{a\dot{b}}\xi^{c}
=\partial_{a\dot{b}}\xi^{c}+\Gamma^{c}_{da\dot{b}}\xi^{d},
\end{equation}
where $\partial_{a\dot{b}}$ are ordinary spinor derivatives, $\Gamma^{c}_{da\dot{b}}$
are spin coefficients.\\
 \hspace*{7.5mm}Now let
 \begin{eqnarray}
 \partial_{0\dot{0}}=l^{\mu}\partial_{\mu}\equiv D,&
 \partial_{1\dot{1}}=n^{\mu}\partial_{\mu}\equiv \Delta,\nonumber\\
 \partial_{0\dot{1}}=m^{\mu}\partial_{\mu}\equiv \delta,
 & \partial_{1\dot{0}}=\bar{m}^{\mu}\partial_{\mu}\equiv\bar{\delta}.
 \end{eqnarray}
Then the Dirac equations (\ref{dirac1}) and (\ref{dirac2}) can be rewritten as four
coupled equations
\begin{eqnarray}\label{dirac3}
&&\left(D+\epsilon-\rho\right)F_{1}+\left(\bar{\delta}+\pi-\alpha\right)F_{2}=0,\nonumber\\
&&\left(\Delta+\mu-\gamma\right)F_{2}+\left(\delta+\beta-\tau\right)F_{1}=0,\nonumber\\
&&\left(D+\epsilon^{\ast}-\rho^{\ast}\right)G_{2}-\left(\bar{\delta}+\pi^{\ast}-\alpha^{\ast}\right)G_{1}=0,\nonumber\\
&&\left(\Delta+\mu^{\ast}-\gamma^{\ast}\right)G_{2}+\left(\delta+\beta^{\ast}-\tau^{\ast}\right)G_{1}=0,
\end{eqnarray}
where $F_{1},F_{2},G_{1},G_{2}$ are four-component spinors with
$F_{1}=P^{0},F_{2}=P^{1},G_{1}=Q^{\dot{1}},G_{2}=Q^{\dot{0}}$.
$\alpha,\beta,\gamma,\epsilon,\mu,\pi,\rho,\tau$ etc. are Newman-Penrose symbols, while
$\alpha^{\ast},\beta^{\ast}$ etc. are, respectively, the complex conjugates of
$\alpha,\beta$ etc.. The Newman-Penrose symbols are
\begin{eqnarray}
&&\rho=l_{\mu;\nu}m^{\mu}\bar{m}^{\nu}=-\frac{1}{\bar{\varrho}^{\ast}},
\;\;\;\;\;\;k=l_{\mu;\nu}m^{\mu}l^{\nu}=0,\;\;\;\;\;\;\lambda=-n_{\mu;\nu}\bar{m}^{\mu}\bar{m}^{\nu}=0,\nonumber\\
&&\tau=l_{\mu;\nu}m^{\mu}n^{\nu}=-\frac{ia\sqrt{\Delta_{\theta}}\sin\theta}{\sqrt{2}\varrho^2},
\;\;\;\;\sigma=l_{\mu;\nu}m^{\mu}m^{\nu}=0,\;\;\;\;\nu=-n_{\mu;\nu}\bar{m}^{\mu}n^{\nu}=0,\nonumber\\
&&\mu=-n_{\mu;\nu}\bar{m}^{\mu}m^{\nu}=-\frac{\Delta_{r}}{2\varrho^{2}\bar{\varrho}^{\ast}},
\;\;\;\;\;\;\;\pi=-n_{\mu;\nu}\bar{m}^{\mu}l^{\nu}=\frac{ia\sqrt{\Delta_{\theta}\sin\theta}}{\sqrt{2}
\left(\bar{\varrho}^{\ast}\right)^{2}},\nonumber\\
&&\epsilon=\frac{1}{2}\left(l_{\mu;\nu}n^{\mu}l^{\nu}-m_{\mu;\nu}\bar{m}^{\mu}l^{\nu}\right)=0,\nonumber\\
&&\gamma=\frac{1}{2}\left(l_{\mu;\nu}n^{\mu}n^{\nu}-m_{\mu;\nu}\bar{m}^{\mu}n^{\nu}\right)
=\frac{1}{4\varrho^{2}}\frac{d\Delta_{r}}{dr}+\mu,\nonumber\\
&&\beta=\frac{1}{2}\left(l_{\mu;\nu}n^{\mu}m^{\nu}-m_{\mu;\nu}\bar{m}^{\mu}m^{\nu}\right)=
\frac{1}{2\sqrt{2}\bar{\varrho}\sin\theta}\frac{d\left(\sqrt{\Delta_{\theta}}\sin\theta\right)}{d\theta},\nonumber\\
&&\alpha=\frac{1}{2}\left(l_{\mu;\nu}n^{\mu}\bar{m}^{\nu}-m_{\mu;\nu}\bar{m}^{\mu}\bar{m}^{\nu}\right)=\pi-\beta^{\ast}.
\end{eqnarray}
\hspace*{7.5mm}Take three transformations as follows:
\begin{eqnarray}
&F_{1}=e^{-i\omega t}e^{im\varphi}f_{1}(r,\theta),\;\;\;F_{2}=e^{-i\omega
t}e^{im\varphi}f_{2}(r,\theta),
\nonumber\\
&G_{1}=e^{-i\omega t}e^{im\varphi}g_{1}(r,\theta),\;\;\;G_{2}=e^{-i\omega
t}e^{im\varphi}g_{2}(r,\theta),\\
&U_{1}(r,\theta)=\bar{\varrho}^{\ast}f_{1}(r,\theta),\;\;\;\;\;\;\;\;U_{2}(r,\theta)=f_{2}(r,\theta),\nonumber\\
&V_{1}(r,\theta)=g_{1}(r,\theta),\;\;\;\;\;\;\;\;\;\;\;V_{2}(r,\theta)=\bar{\varrho}g_{2}(r,\theta),\\
&U_{1}=R_{-\frac{1}{2}}(r)S_{-\frac{1}{2}}(\theta),\;\;\;\;\;\;\;\;U_{2}=R_{+\frac{1}{2}}(r)S_{+\frac{1}{2}}(\theta)
,\nonumber\\
&V_{1}=R_{+\frac{1}{2}}(r)S_{-\frac{1}{2}}(\theta),\;\;\;\;\;\;\;\;V_{2}=R_{-\frac{1}{2}}(r)S_{+\frac{1}{2}}(\theta).
\end{eqnarray}
Eqs.(\ref{dirac3}) turn into
\begin{eqnarray}\label{dirac4}
&&{\cal{D}}_{0}R_{-\frac{1}{2}}S_{-\frac{1}{2}}+\sqrt{\frac{\Delta_{\theta}}{2}}{\cal{L}}_{\frac{1}{2}}
R_{+\frac{1}{2}}S_{+\frac{1}{2}}=0,\nonumber\\
&&\Delta_{r}{\cal{D}}^{+}_{\frac{1}{2}}R_{+\frac{1}{2}}S_{+\frac{1}{2}}-\sqrt{2\Delta_{\theta}}
{\cal{L}}^{+}_{\frac{1}{2}}R_{-\frac{1}{2}}S_{-\frac{1}{2}}=0,\nonumber\\
&&{\cal{D}}_{0}R_{-\frac{1}{2}}S_{+\frac{1}{2}}-\sqrt{\frac{\Delta_{\theta}}{2}}
{\cal{L}}^{+}_{\frac{1}{2}}R_{+\frac{1}{2}}S_{-\frac{1}{2}}=0,\nonumber\\
&&\Delta_{r}{\cal{D}}^{+}_{\frac{1}{2}}R_{+\frac{1}{2}}S_{-\frac{1}{2}}+\sqrt{2\Delta_{\theta}}
{\cal{L}}^{+}_{\frac{1}{2}}R_{-\frac{1}{2}}S_{+\frac{1}{2}}=0,
\end{eqnarray}
where
\begin{eqnarray}
&&{\cal{D}}_{s}=\partial_{r}-\frac{i\Sigma
K}{\Delta_{r}}+\frac{s}{\Delta_{r}}\frac{d\Delta_{r}}{dr},\nonumber\\
&&{\cal{D}}^{+}_{s}=\partial_{r}+\frac{i\Sigma
K}{\Delta_{r}}+\frac{s}{\Delta_{r}}\frac{d\Delta_{r}}{dr},\nonumber\\
&&{\cal{L}}_{s}=\partial_{\theta}-\frac{\Sigma
H}{\Delta_{\theta}}+\frac{s}{\sqrt{\Delta_{\theta}}\sin\theta}\frac{d
\left(\sqrt{\Delta_{\theta}}\sin\theta\right)}{d\theta},\nonumber\\
&&{\cal{L}}^{+}_{s}=\partial_{\theta}+\frac{\Sigma
H}{\Delta_{\theta}}+\frac{s}{\sqrt{\Delta_{\theta}}\sin\theta}\frac{d
\left(\sqrt{\Delta_{\theta}}\sin\theta\right)}{d\theta},\nonumber\\
&&K=\left(r^{2}+a^{2}\right)\omega-am,\nonumber\\
&&H=a\omega\sin\theta-\frac{m}{\sin\theta}.
\end{eqnarray}
By using separation of variables, Eqs.(\ref{dirac4}) become
\begin{eqnarray}\label{dirac5}
&&{\cal{D}}_{0}R_{-\frac{1}{2}}=\lambda_{1}R_{+\frac{1}{2}},\;\;\;\;\;\;\;\;\;
\sqrt{\frac{\Delta_{\theta}}{2}}{\cal{L}}_{\frac{1}{2}}S_{+\frac{1}{2}}=-\lambda_{1}S_{-\frac{1}{2}},\nonumber\\
&&\Delta_{r}{\cal{D}}^{+}_{\frac{1}{2}}R_{+\frac{1}{2}}=\lambda_{2}R_{-\frac{1}{2}},\;\;\;
\sqrt{2\Delta_{\theta}}{\cal{L}}^{+}_{\frac{1}{2}}S_{-\frac{1}{2}}=\lambda_{2}S_{+\frac{1}{2}},\nonumber\\
&&{\cal{D}}_{0}R_{-\frac{1}{2}}=\lambda_{3}R_{+\frac{1}{2}},\;\;\;\;\;\;\;\;\;
\sqrt{\frac{\Delta_{\theta}}{2}}{\cal{L}}^{+}_{\frac{1}{2}}S_{-\frac{1}{2}}=\lambda_{3}S_{+\frac{1}{2}},\nonumber\\
&&\Delta_{r}{\cal{D}}^{+}_{\frac{1}{2}}R_{+\frac{1}{2}}=\lambda_{4}R_{-\frac{1}{2}},\;\;\;
\sqrt{2\Delta_{\theta}}{\cal{L}}_{\frac{1}{2}}S_{+\frac{1}{2}}=-\lambda_{4}S_{-\frac{1}{2}},
\end{eqnarray}
where $\lambda_{1},\lambda_{2},\lambda_{3},\lambda_{4}$ are the separation constants,
and
\begin{equation}
\lambda_{1}=\lambda_{3}=\frac{1}{2}\lambda_{2}=\frac{1}{2}\lambda_{4}\equiv\lambda.
\end{equation}
Then Eqs.(\ref{dirac5}) can be written as
\begin{eqnarray}\label{dirac6}
&&{\cal{D}}_{0}R_{-\frac{1}{2}}=\lambda R_{+\frac{1}{2}},\;\;\;\;\;\;\;\;\;\;\;\;\;\;\;
\Delta_{r}{\cal{D}}^{+}_{\frac{1}{2}}R_{+\frac{1}{2}}=2\lambda
R_{-\frac{1}{2}},\nonumber\\
&&\Delta^{\frac{1}{2}}_{\theta}{\cal{L}}_{\frac{1}{2}}S_{+\frac{1}{2}}=-\sqrt{2}\lambda
S_{-\frac{1}{2}},\;\;\;\Delta^{\frac{1}{2}}_{\theta}{\cal{L}}^{+}_{\frac{1}{2}}S_{-\frac{1}{2}}=\sqrt{2}\lambda
S_{+\frac{1}{2}},
\end{eqnarray}
\hspace*{7.5mm}Substituted $\sqrt{2}\lambda$ with $\lambda$, $\sqrt{2}R_{-\frac{1}{2}}$
with $R_{-\frac{1}{2}}$, Eqs.(\ref{dirac6}) can be written as
\begin{eqnarray}\label{dirac7}
&&\Delta^{\frac{1}{2}}_{r}{\cal{D}}_{0}R_{-\frac{1}{2}}=\lambda\Delta^{\frac{1}{2}}_{r}R_{+\frac{1}{2}},\;\;\;
\Delta^{\frac{1}{2}}_{r}{\cal{D}}^{+}_{0}\Delta^{\frac{1}{2}}_{r}R_{+\frac{1}{2}}=\lambda
R_{-
\frac{1}{2}},\nonumber\\
&&\Delta^{\frac{1}{2}}_{\theta}{\cal{L}}_{\frac{1}{2}}S_{+\frac{1}{2}}=-\lambda
S_{-\frac{1}{2}},\;\;\;\;\;\;\;\;\;\;\;\;
\Delta^{\frac{1}{2}}_{\theta}{\cal{L}}^{+}_{\frac{1}{2}}S_{-\frac{1}{2}}=\lambda S_{+
\frac{1}{2}}.
\end{eqnarray}
Set
\begin{equation}
\Delta^{\frac{1}{2}}_{r}R_{+\frac{1}{2}}=P_{+\frac{1}{2}},\;\;\;\;\;\;R_{-\frac{1}{2}}=P_{-\frac{1}{2}},
\end{equation}
Eqs.(\ref{dirac7}) reduce to
\begin{eqnarray}
\label{radial}&&\Delta^{\frac{1}{2}}_{r}{\cal{D}}^{+}_{0}P_{+\frac{1}{2}}=\lambda
P_{-\frac{1}{2}},
\;\;\;\;\;\;\Delta^{\frac{1}{2}}_{r}{\cal{D}}_{0}P_{-\frac{1}{2}}=\lambda
P_{+\frac{1}{2}},\\
\label{angular1}&&\Delta^{\frac{1}{2}}_{\theta}{\cal{L}}_{\frac{1}{2}}S_{+\frac{1}{2}}=-\lambda
S_{-\frac{1}{2}},\;\;\;\;\Delta^{\frac{1}{2}}_{\theta}{\cal{L}}^{+}_{\frac{1}{2}}S_{-\frac{1}{2}}=\lambda
S_{+\frac{1}{2}}.
\end{eqnarray}
\hspace*{7.5mm}Introducing the tortoise coordinate transformation from the radial
variable $r$ to the tortoise coordinate $r_{\ast}$ which is given by
\begin{equation}
\frac{d}{dr_{\ast}}=\frac{\Delta_{r}d}{\bar{\omega}^{2}dr},
\end{equation}
where
\begin{equation}
\bar{\omega}^2=r^{2}+a^{2}-\frac{am}{\omega}.
\end{equation}
We set
\begin{equation}
{\cal{D}}_{0}=\frac{\bar{\omega}^{2}}{\Delta_{r}}\left(\frac{d}{dr_{\ast}}+i\sigma\right),
\end{equation}
where
\begin{equation}
\sigma=-\Sigma\omega,
\end{equation}
and
\begin{equation}
{\cal{D}}^{+}_{0}=\frac{\bar{\omega}^{2}}{\Delta_{r}}\left(\frac{d}{dr_{\ast}}-i\sigma\right).
\end{equation}
Eqs.(\ref{radial}) is reduced to
\begin{eqnarray}
\label{radial1-1}\left(\frac{d}{dr_{\ast}}-i\sigma\right)P_{+\frac{1}{2}}=\lambda\frac{\Delta^{\frac{1}{2}}_{r}}
{\bar{\omega}^{2}}P_{-\frac{1}{2}},\\
\label{radial1-2}\left(\frac{d}{dr_{\ast}}+i\sigma\right)P_{-\frac{1}{2}}=\lambda\frac{\Delta^{\frac{1}{2}}_{r}}
{\bar{\omega}^{2}}P_{+\frac{1}{2}}.
\end{eqnarray}
By setting
\begin{equation}
Z_{\pm}=P_{+\frac{1}{2}}\pm P_{-\frac{1}{2}},
\end{equation}
eq.(\ref{radial1-1}) and (\ref{radial1-2}) changed to be
\begin{eqnarray}
\label{radial2-1}\left(\frac{d}{dr_{\ast}}-\lambda\frac{\Delta^{\frac{1}{2}}_{r}}{\bar{\omega}^{2}}\right)Z_{+}=i\sigma
Z_{-},\\
\label{radial2-2}\left(\frac{d}{dr_{\ast}}+\lambda\frac{\Delta^{\frac{1}{2}}_{r}}{\bar{\omega}^{2}}\right)Z_{-}=i\sigma
Z_{+}.
\end{eqnarray}
\hspace*{7.5mm}From eq.(\ref{radial2-1}) and (\ref{radial2-2}), we obtain the radial
wave equation
\begin{equation}\label{vexp}
\left(\frac{d^{2}}{dr^{2}_{\ast}}+\sigma^{2}\right)Z_{\pm}=V_{\pm}Z_{\pm},
\end{equation}
where
\begin{eqnarray}\label{wexp}
V_{\pm}&=&\lambda^{2}W^{2}\pm\lambda\frac{dW}{dr_{\ast}}\nonumber\\
W&=&\frac{\Delta^{\frac{1}{2}}_{r}} {\bar{\omega}^{2}}
\end{eqnarray}
$\lambda$ in Eqs.(\ref{dirac7})-(\ref{wexp}) is a separation constant and $\sigma$ is
the QNMs of the black hole. We derived the expression of $\lambda$ at the limit of
small cosmological constant $\Lambda$ and slow rotating black hole in Appendix
\ref{eqr}. It can be written as a function of
 $l$ (or $j$), $m$, $a$ and $\sigma$. \\
\psfrag{Lambda}{$\Lambda$}\psfrag{sigma}{$\sigma$}
\section{PROPERTIES OF MASSLESS DIRAC FIELD EFFECTIVE POTENTIAL}
\hspace*{7.5mm}From the Schr\"{o}dinger-like equations in eq.(\ref{vexp}), we can
evaluate the QNMs. The form of $V_{+}$ and $V_{-}$ shown in eq.(\ref{wexp}) are
super-symmetric partners derived from the same super-potential $W$
\cite{superpotential}. Ref \cite{sameV} has proved that potentials related in this ways
have the same spectral of QNMs. Thus we deal with eq.(\ref{vexp}) with potential
$V_{+}$ in evaluating the QNMs. The effective potential also depends on $\sigma$.\\
\begin{figure}[htbp]
\centerline{\includegraphics{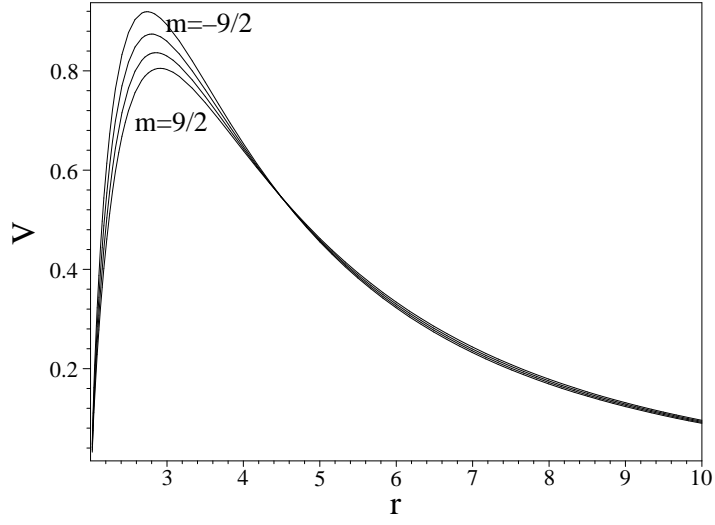}} \caption{Variation of the effective potential for
massless neutrino with
$\Lambda=0.01,\sigma=1,a=0.1,Q=0.1,l=5,j=\frac{9}{2},m=\frac{9}{2},\frac{3}{2},-\frac{3}{2},-\frac{9}{2}$.}
\label{mv}
\end{figure}\\

\begin{figure}[htbp]
\centerline{\includegraphics{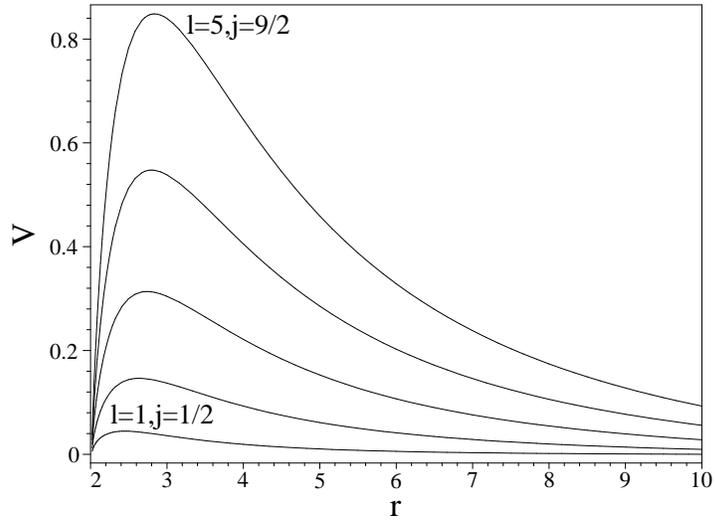}} \caption{Variation of the effective potential for
massless neutrino with
$\Lambda=0.01,\sigma=1,a=0.1,Q=0.1,m=\frac{1}{2},l=1,2,3,4,5,j=l-1/2$.} \label{lv}
\end{figure}

\begin{figure}[htbp]
\centerline{\includegraphics{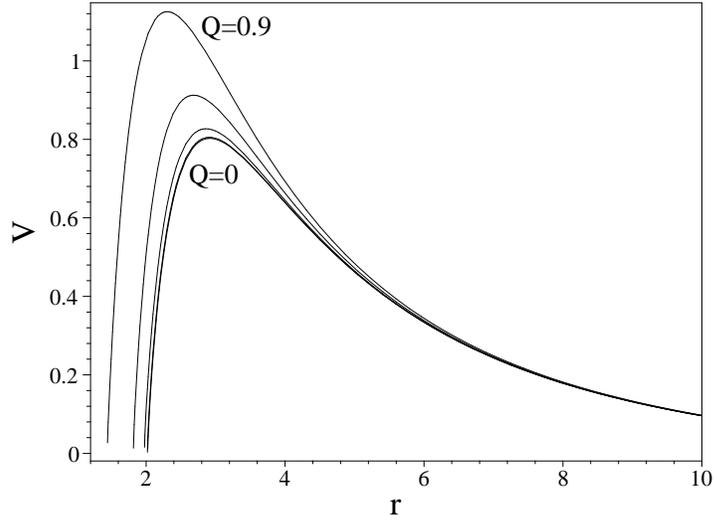}} \caption{Variation of the effective potential for
massless neutrino with
$\Lambda=0.01,\sigma=1,a=0.1,m=\frac{9}{2},l=5,j=\frac{9}{2},Q=0,0.1,0.3,0.6,0.9$.} \label{qv}
\end{figure}

\begin{figure}[htbp]
\centerline{\includegraphics{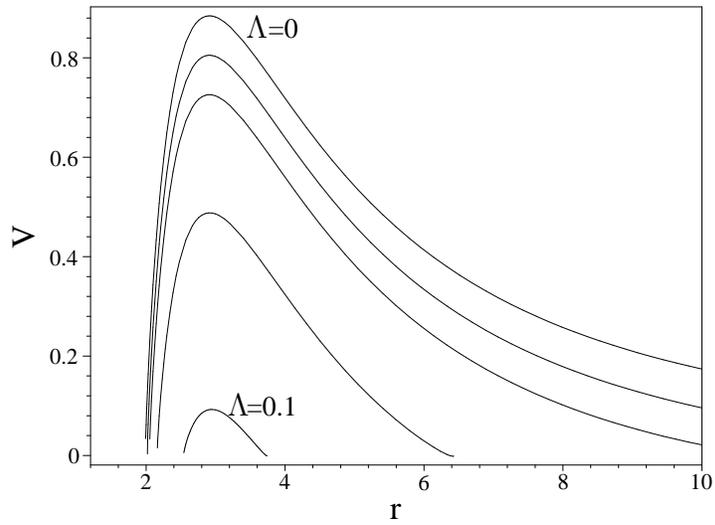}} \caption{Variation of the effective potential
for massless neutrino with
$\sigma=1,a=0.1,m=\frac{9}{2},l=5,j=\frac{9}{2},Q=0.1,\Lambda=0,0.01,0.02,0.05,0.1$.}
\label{lav}
\end{figure}

\begin{figure}[htbp]
\centerline{\includegraphics{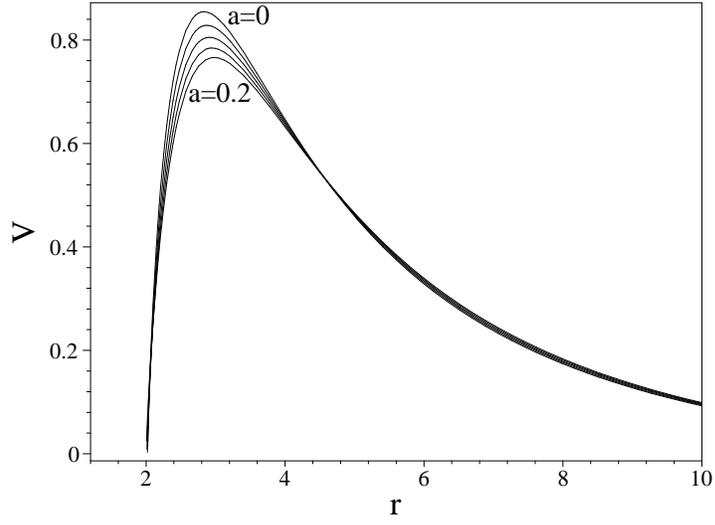}} \caption{Variation of the effective potential for
massless neutrino with
$\Lambda=0.01,\sigma=1,Q=0.1,m=\frac{9}{2},l=5,j=\frac{9}{2},a=0,0.05,0.1,0.15,0.2$.}
\label{av}\end{figure}

\begin{figure}[htbp]
\centerline{\includegraphics{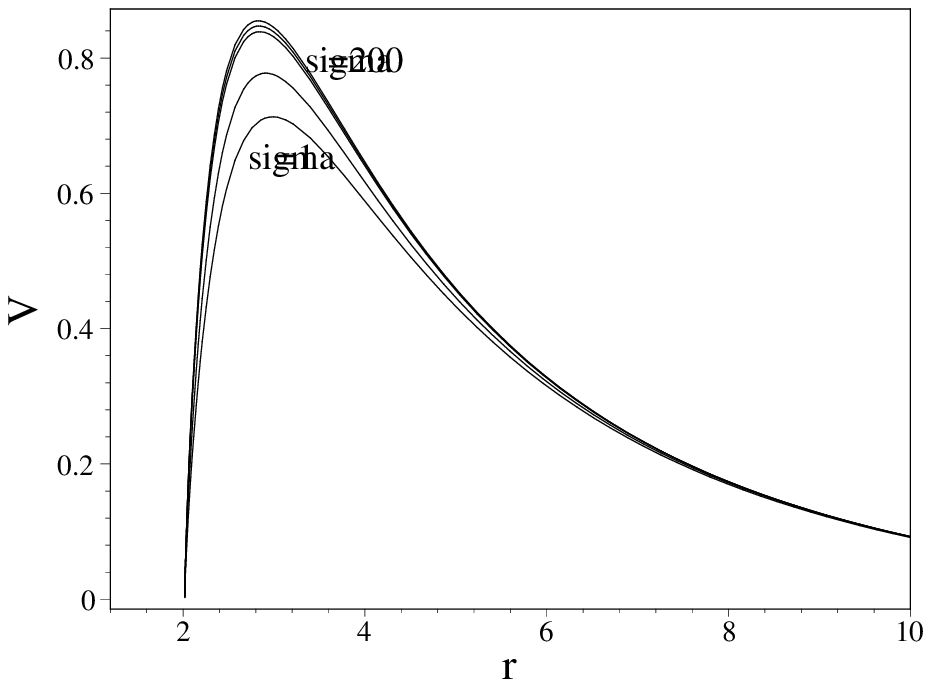}} \caption{Variation of the effective potential for
massless neutrino with
$\Lambda=0.01,a=0.1,Q=0.1,m=\frac{9}{2},l=5,j=\frac{9}{2},\omega=1,10,20,100,200$.}
\label{ov}
\end{figure}

The QNMs are decided by the effective potential. Here we analyze the dependence of the
effective potential on parameters $m$, $l$, $Q$, $\Lambda$,$a$ and $\sigma$. The
effective potential as a function of $r$ is plotted for some configurations of $m$,
$l$, $Q$, $\Lambda$,$a$ and $\sigma$ in Fig.\ref{mv}-\ref{ov}. From these figures, we
can see that the dependence of $V$ on $l$, $\Lambda$, $Q$ is stronger
 than on $m$, $a$ and $\sigma$. Because there exists cosmological a constant $\Lambda$, the space time
possesses two horizons: the black hole horizon $r=r_{e}$ and the cosmological horizon
$r=r_{c}$. While $r$ varies from $r_{e}$ to $r_{c}$, the effective potential $V$
reduces to zero. \\
\hspace*{7.5mm}Because of the rotation of the Kerr black hole, the azimuthal degeneracy
of magnetic quantum number $m$ is destroyed. On the Fig.\ref{mv}, we show the
dependence of effective potential on $m$, and find that negative $m$ will increase the
maximum values of
the effective potential and decrease the position of the peak.\\
\hspace*{7.5mm}In the Fig.\ref{lv}, we show the dependence of the effective potential
on angular momentum number $l$. It is clear the peak value and position of the
potential increase with $l$. \\
\hspace*{7.5mm}Fig.\ref{qv} shows the dependence of the effective potential on the
electric charge $Q$ of black hole. The electric charge of black hole will increase the
the peak value, decrease the position of the peak and change the position of black hole horizon.\\
 \hspace*{7.5mm}Fig.\ref{lav} shows the dependence of effective potential on the cosmological
  constant $\Lambda$. Increasing of $\Lambda$ reduces the peak value of the effective potential,
  decreases the cosmological horizon radius $r_{c}$ and increases the black hole horizon radius $r_{e}$. \\
\hspace*{7.5mm}In the Fig.\ref{av}, we show the dependence of the effective potential
on $a$. For the positive value of $m$, rotation of the black hole reduce the peak value
and increase the
 position of the peak.\\
 \hspace*{7.5mm}For the limit of small $a$ and the perturbation method used in Appendix
 \ref{eqr}, the effective of rotation to the separation constant $\lambda$ is small.
 When we consider the dependence of the effective potential on $\sigma$, we neglect the
change of $\lambda$. For $\lambda=5$, the Fig.\ref{ov} suggests the potential varies
slowly as $\sigma$ increases and approaches a limiting position.\\
\section{MASSLESS NEUTRINO QNMS OF A KERR-NEWMAN-DE SITTER BLACK HOLE}
\begin{figure}[htbp]
\centerline{\includegraphics{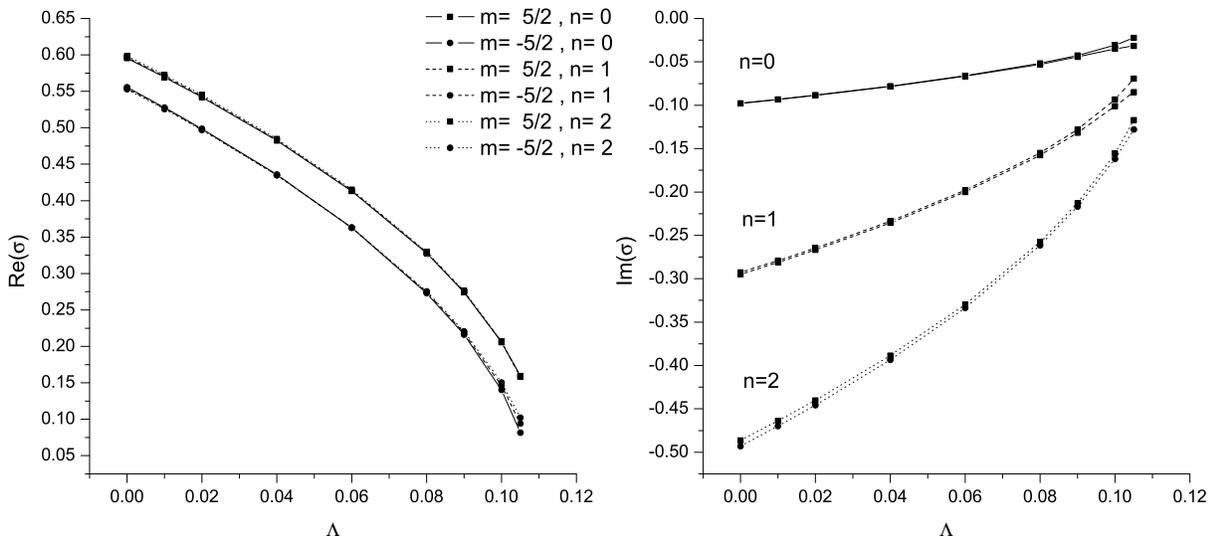}} \caption{Real and imaginary parts of the
massless neutrino QNMs as a function of $\Lambda$ for
$Q=0.1,a=0.1,l=3,j=\frac{5}{2},m=\pm\frac{5}{2}$.} \label{qql}
\end{figure}

\setcounter{equation}{0}\hs In this section, we evaluate the QNMs by using
P\"{o}shl-Teller potential approximation instead of more popular WKB approximation.
Konoplya \cite{konoplya} used sixth-order WKB approximation to calculated the QNMs of a
D-dimensional Schwarzshild black hole and compared with the result of third-order WKB
approximation. Zhidenko \cite{zhidenko} calculated QNMs of a Schwarzschild-de Sitter
black hole by using sixth-order WKB approximation and the approximation by
P\"{o}shl-Teller potential. The results of Ref \cite{konoplya} and \cite{zhidenko} show
that for large $l$ and $\Lambda$ P\"{o}shl-Teller potential approximation can give
results
agree well with the sixth-order WKB approximation.\\
\hspace*{7.5mm}Proposed by Ferrari and Mashhoon \cite{ferrari}, the P\"{o}shl-Teller
potential approximation method use P\"{o}shl-Teller approximate potential
\begin{equation}
V_{PT}=\frac{V_{0}}{\cosh^{2}\left(r^{\ast}/b\right)}.
\end{equation}
It contains two free parameters ($V_{0}$and $b$) which are used to fit the height and
the second derivative of the potential $V(r^{\ast})$ at the maximum
\begin{equation}
\frac{1}{b^{2}}=-\frac{1}{2V_{0}}\left[\frac{d^{2}V}{dr^{2}}\right]_{r\rightarrow
r_{0}}.
\end{equation}
\hspace*{7.5mm}The QNMs of the P\"{o}shl-Teller potential can be evaluated
analytically:
\begin{equation}\label{pt}
\sigma=\frac{1}{b}\left[\sqrt{V_{0}b^2-\frac{1}{4}}-\left(n+\frac{1}{4}\right)i\right],
\end{equation}
where $n$ is the mode number and $n<l$ for low-laying modes. From eq.(\ref{pt}), the
real parts of the QNMs are
 independent of $n$. This relates to how to approximates the effective potential. \\
 \hspace*{7.5mm}The effective potential also depends on $\sigma$. This will complicate matters in
eq.(\ref{pt}) because there are $\sigma$ dependence on both sides of the equation. Thus
we cannot obtain $\sigma$ directly. For slowly rotating black hole with little
cosmological constant $\Lambda$, we can expand separation constant $\lambda$ as
series of $a\ll1$, and expand the effective potential in power series of $a$.\\
\hspace*{7.5mm}Firstly we express the position of the peak of the effective potential
as series up to order $a^{5}$,
\begin{eqnarray}
r_{max}&=&r_{0}+r_{1}a+r_{2}a^{2}+r_{3}a^{3}+r_{4}a^{4}+r_{5}a^{5}\nonumber\\
&=&r_{0}+\Sigma_{0},
\end{eqnarray}
and
\begin{eqnarray}
0=V'\left(r_{max}\right)&=&V'\left(r_{0}\right)+\Sigma_{0} V''
\left(r_{0}\right)+\frac{1}{2}\Sigma_{0}^{2}V'''\left(r_{0}\right)\nonumber\\
 &&+\frac{1}{6}\Sigma_{0}^{3}V^{(4)}\left(r_{0}\right)+\frac{1}{24}\Sigma_{0}
 ^{4}V^{(5)}\left(r_{0}\right)+\frac{1}{120}\Sigma_{0}^{5}V^{(6)}\left(r_{0}\right),\nonumber\\
\end{eqnarray}
where $r_{0}$ is the position of the peak of effective potential for the non-rotating
black hole. Eq.(\ref{wexp}) for the non-rotating black hole case, show that the
expression of effective potential $V$ does not depend on $\sigma$ and can be solved
independently. We evaluate the coefficients $r_{i}$'s order by solving this equation.
The expression of $r_{max}$ contains $a$ and unknown $\sigma$. We also expand $\sigma$
as $\sigma=\sigma_{0}+\sigma_{1}a+\sigma_{2}a^{2}+\sigma_{3}a^{3}+\sigma_{4}a^{4}
+\sigma_{5}a^{5}$ and plug in the expansion for $r_{max}$, and then expand the
derivation of the potential $V_{0}^{(n)}$ performed with respect to $r_{\ast}$. We plug
all these expansions back to eq.(\ref{pt}) and solve the coefficients $\sigma_{i}$'s
self-consistently order by order in $a$.
\begin{figure}[htbp]
\centerline{\includegraphics{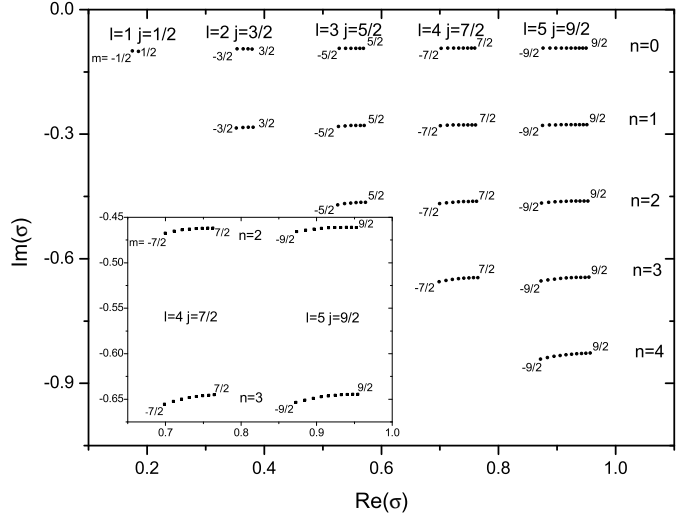}} \caption{QNMs of massless neutrino for
$\Lambda=0.01,Q=0.1,a=0.1$.} \label{qlm}
\end{figure}\\
\hspace*{7.5mm}Now we evaluate massless neutrino QNMs of a Kerr-Newman-de Sitter black
hole by using P\"{o}shl-Teller potential approximation for $\Lambda=0.01$, $Q=0.1$, and
$a=0.1$, plot the results in Fig.\ref{qlm} and list them in Appendix \ref{qml}. The
results show that the real parts of the QNMs increase with $l$ and the magnitude of the
imaginary parts increase with $n$.
 \begin{figure}[htbp]
\centerline{\includegraphics{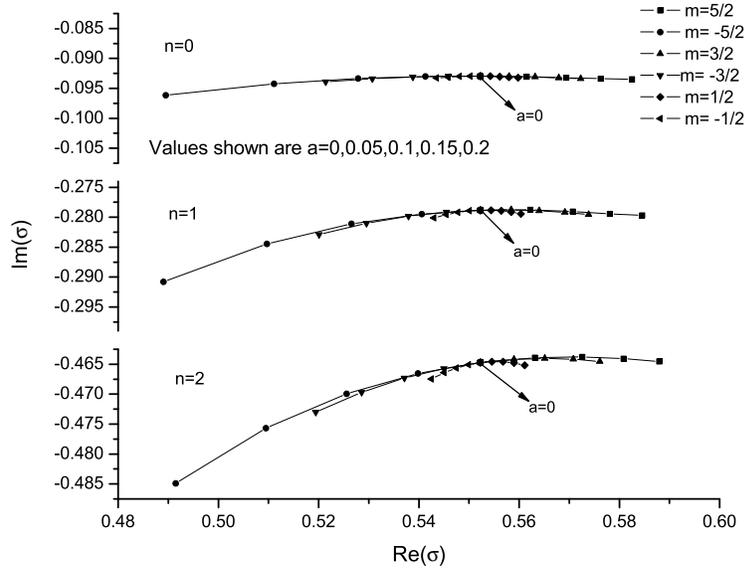}} \caption{QNMs of massless neutrino vary with $a$
for
$\Lambda=0.01,Q=0.1,l=3,j=\frac{5}{2},m=\pm\frac{5}{2},\pm\frac{3}{2},\pm\frac{1}{2}$.}
\label{qa}\end{figure}
\begin{figure}[htbp]
\centerline{\includegraphics{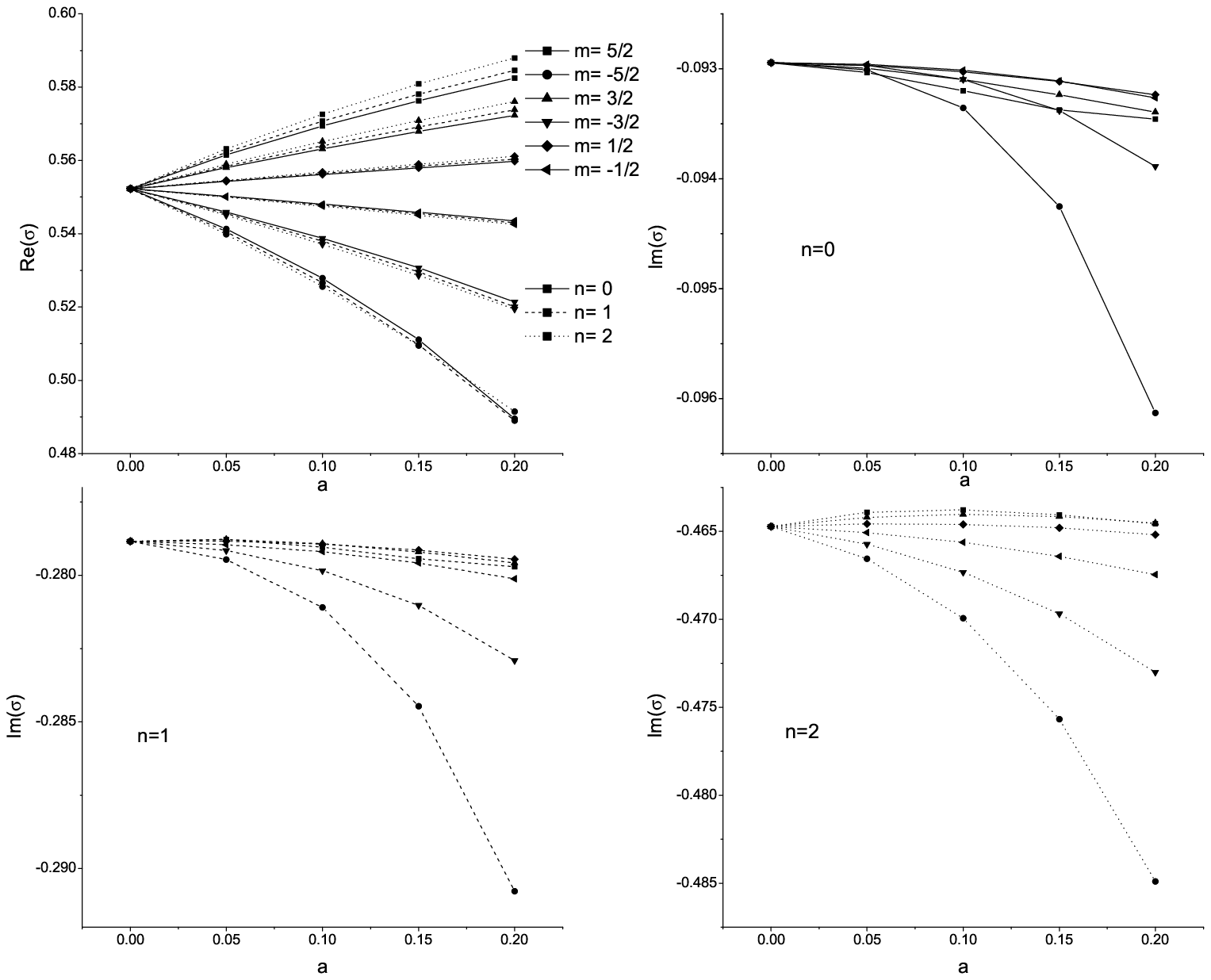}} \caption{Real and imaginary parts of the
massless neutrino QNMs as a function of $a$ for
$\Lambda=0.01,Q=0.1,l=3,j=\frac{5}{2},m=\pm\frac{5}{2},\pm\frac{3}{2},\pm\frac{1}{2}$.}
\label{qqa}
\end{figure}\\
\hspace*{7.5mm}Figures of QNMs varying with $a$ is plotted  in the Fig.\ref{qa} and
plot the real and imaginary parts of QNMs as a function of $a$ on the Fig.\ref{qqa}. We
vary $a$ from $0$ to 0.2 to satisfy the condition $a\ll1$. Because of the spherical
symmetry of a non-rotating black hole, the QNMs is the azimuthal degenerate which can
be verified in Fig.\ref{qa} and Fig.\ref{qqa}. $a=0$ means non-rotating case, which is
Reissner-Nordstr\"{o}m-de Sitter black hole here. Different values of $m$ have the same
QNMs in the Fig.\ref{qa} and Fig.\ref{qqa}. They also clearly display the split of
azimuthal degeneracy as $a$ increasing from $0$ to $0.2$.\\
 \hspace*{7.5mm} From the Fig.\ref{qqa}, we see that for a rotating black hole, the real
 parts of QNMs is also related to the $n$ though we use the P\"{o}shl-Teller potential
 approximation. For a slowly rotating Kerr-Newman-de Sitter black hole with little
 value of cosmological constant $\Lambda$, the separation constant $\lambda$ can be
 written as eq.(\ref{lambda}) and the function of $\sigma$ which relates to $n$. As to the increasing of
 $a$, the real parts of QNMs increase for positive $m$ and decrease for negative $m$,
 while the split of real parts for negative $m$ is bigger than the positive case with the same magnitude of $m$.
  The split of real parts for $n$ increase with $a$ and the magnitude of $m$. $n$ increases the real
  parts for the positive $m$ and decreases the imaginary parts.  The real parts split of $n$ for positive $m$
  is bigger than imaginary case. For the negative $m$, the larger magnitude of  $m$ change the imaginary parts more.
   The rotation increases the magnitude of imaginary parts and the split of different values of $m$.
   For the positive $m$ case, it is more complex but the Fig.\ref{qqa} shows that the imaginary parts of different
    positive $m$ for different $n$ trend to the same values and the tendency is more clearly for large $n$. \\
  \begin{figure}[htbp]
\centerline{\includegraphics{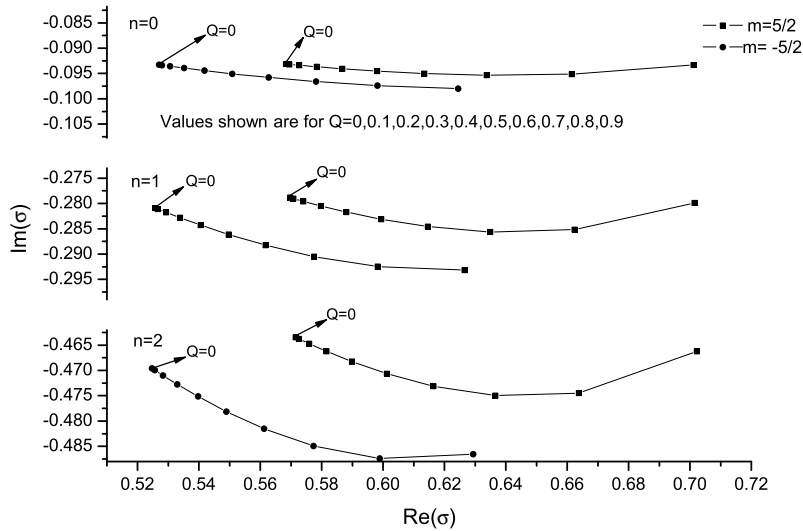}} \caption{QNMs of massless neutrino vary with $Q$
for $\Lambda=0.01,a=0.1,l=3,j=\frac{5}{2},m=\pm\frac{5}{2}$.} \label{qq}\end{figure}
\begin{figure}[htbp]
\centerline{\includegraphics{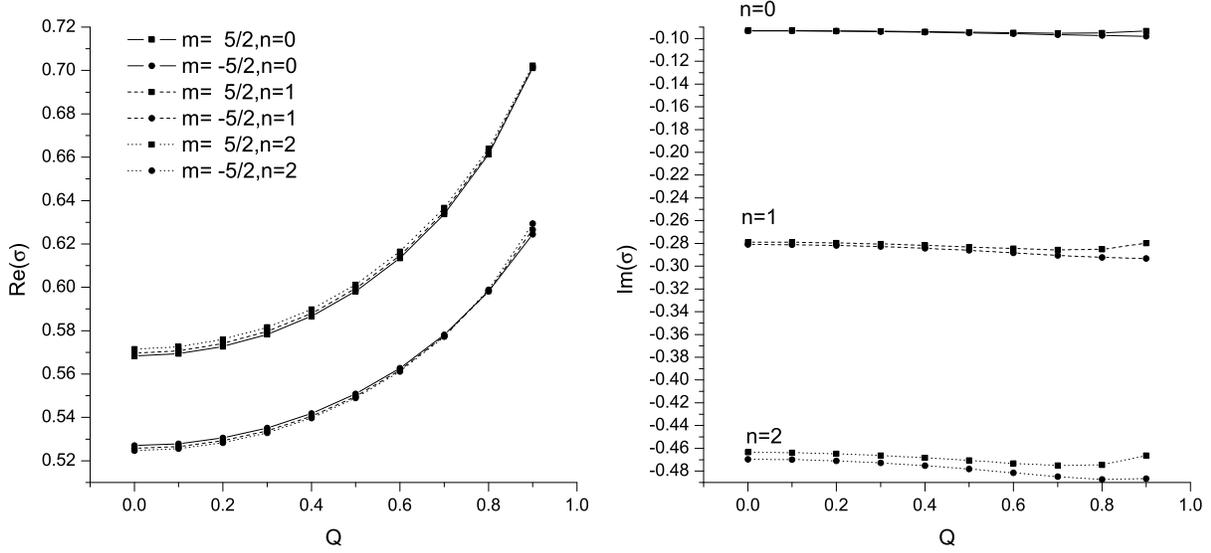}} \caption{Real and imaginary parts of the
massless neutrino QNMs as a function of $Q$ for
$\Lambda=0.01,a=0.1,l=3,j=\frac{5}{2},m=\pm\frac{5}{2}$.} \label{qqq}
\end{figure}
\hspace*{7.5mm}We plot the image of QNMs varying with $Q$ in the Fig.\ref{qq} and plot
the real and imaginary parts of QNMs as a function of $Q$ on the Fig.\ref{qqq}. The
real parts of QNMs increase with $Q$, the split of different values of $n$ decrease
first and increase later. For example, while $m=-\frac{5}{2}$ the real parts of $n=0$
is larger than $n=2$ for $Q=0$ and the other way round for $Q=0.9$. The split of
imaginary
parts of different $m$ for the same $n$ increase with $Q$.\\
\begin{figure}[htbp]
\centerline{\includegraphics{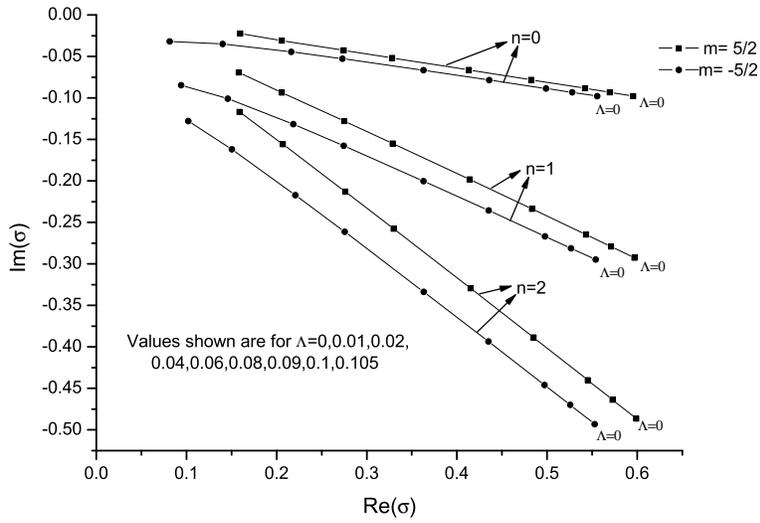}} \caption{QNMs of massless neutrino vary with
$\Lambda$ for $Q=0.1,a=0.1,l=3,j=\frac{5}{2},m=\pm\frac{5}{2}$.} \label{ql}\end{figure}
\hspace*{7.5mm}In Fig.\ref{ql} and Fig.\ref{qql} we plot image of QNMs varying with
cosmological constant $\Lambda$  and the real and imaginary parts of QNMs as a function
of $\Lambda$. Fig.\ref{lav} shows that $\Lambda$ influence the effective potential more
than other parameters. The real parts of QNMs and the magnitude of imaginary parts
decrease with $\Lambda$. Just like varying with $Q$ the split of different values of
$n$ decrease first and increase later. The split of imaginary parts of different $m$
for the same $n$ increase with $Q$. For a sufficient large $\Lambda$, the imaginary
parts split of different $m$ will be larger than that of different $n$, which means
that in the limit of the near extreme $\Lambda$ term for a slowly rotating black hole,
the imaginary parts of the QNMs are mostly determined by $m$ rather than $n$. This is
why in Fig.\ref{ql} the lines of $m=\frac{5}{2}$ and $m=-\frac{5}{2}$ seem to
approach two dots for different $m$.\\
\section{CONCLUSIONS AND DISCUSSIONS}
\setcounter{equation}{0}\hs We have evaluated low-laying massless neutrino QNMs of a
Kerr-Newman-de Sitter black hole by using P\"{o}shl-Teller potential approximation. We
adopt a further approximation by making perturbative expansions for all the quantities
in powers of parameter $a$. \\
\hspace*{7.5mm}In general, the real parts of QNMs increase with $l$. The magnitude of
the imaginary parts increase with $n$. A character feature of the QNMs of rotating
black holes is the split of the azimuthal degeneracy for different values $m$. This is
clearly displayed in Figs.\ref{qlm}, \ref{qa}, \ref{qqa}. For a Kerr-Newman-de Sitter
black hole, massless neutrino perturbation of large $\Lambda$, positive $m$ and small
value of $n$ will
decay slowly.\\
\hspace*{7.5mm}We can expand this methods to others black holes. All these works will
enrich our knowledge about QNMs of different kind of black holes and give direct
identification to distinguish the kind of black holes, through gravitational wave detectors in future.\\
\section*{Acknowledgements}

 \hspace*{7.5mm}The work has been supported by the National Natural Science
Foundation of China (Grant No. 10273017). J. F. Chang thanks Dr. Xian-Hui Ge for his
zealous help during the work.\\

\appendix
\section{ANGULAR EIGENFUNCTIONS AND EIGENVALUES}\label{eqr} \setcounter{equation}{0} \hs The
angular equations (\ref{angular1}) can be combined into
\begin{equation}\label{angular2}
\Delta_{\theta}^{\frac{1}{2}}{\cal{L}}_{\frac{1}{2}}\left[\Delta_{\theta}^{\frac{1}{2}}{\cal{L}}_{\frac{1}{2}}^{+}
S_{-\frac{1}{2}}\right]+\lambda^{2}S_{-\frac{1}{2}}=0,
\end{equation}
and a similar equation for $S_{+\frac{1}{2}}$ obtained by replacing $\theta$ by
$\pi-\theta$. For small value of cosmological constant $\Lambda$, we expand
eq.(\ref{angular2}) as series of $\Lambda$, say by first-order expansion and
\begin{equation}\label{angular3}
\left[{\cal{D}}_{0}+{\cal{D}}_{1}+\left(\frac{1}{3}a^{2}\cos^{2}\theta{\cal{D}}_{0}+{\cal{D}}_{2}\right)\Lambda
+\lambda^{2}\right]S_{-\frac{1}{2}}=0,
\end{equation}
where
\begin{eqnarray}
{\cal{D}}_{0}&\equiv&\frac{d^{2}}{d\theta^{2}}+\cot\theta\frac{d}{d\theta}-\frac{\left(m-
\frac{1}{2}\cos^{2}\theta\right)^{2}}{\sin^{2}\theta}-\frac{1}{2},\nonumber\\
{\cal{D}}_{1}&\equiv&2a\omega
m-a\omega\cos\theta-a^{2}\omega^{2}\sin^{2}\theta,\nonumber\\
{\cal{D}}_{2}&\equiv&-\frac{2}{3}a^{2}\cos\theta\sin\theta\frac{d}{d\theta}-\frac{2}{3}a^{2}m^2
-\frac{1}{3}a^{3}\omega\cos\theta\nonumber\\&&-\frac{2}{3}a^{4}\omega^{2}\sin^{2}\theta+\frac{4}{3}a^{3}\omega
m+\frac{1}{6}a^{2}\sin^{2}\theta\nonumber\\&&-\frac{2}{3}a^{3}\omega
m\cos^{2}\theta+\frac{1}{3}a^{4}\omega^{2}\sin^{2}\theta\cos^{2}\theta\nonumber\\&&
-\frac{1}{3}a^{3}\omega\cos\theta\sin^{2}\theta-\frac{1}{2}a^{2}\cos^{2}\theta.
\end{eqnarray}
The operator ${\cal{D}}_{0}$ has no dependence on $a$ and the solution for
$S_{-\frac{1}{2}}$ of equation
\begin{equation}
{\cal{D}}_{0}S_{-\frac{1}{2}}=-E^{2}S_{-\frac{1}{2}}
\end{equation}
can be written in terms of the standard spin-weighted spherical harmonics (Newman and
Penrose 1966 \cite{newman}; Goldberg \emph{et al}. 1967 \cite{goldberg})
\begin{equation}
S_{-\frac{1}{2}}\left(\cos\theta\right)e^{im\varphi}=\;_{-\frac{1}{2}}Y_{jm}\left(\theta,\varphi\right),\;\;\;\;\;\
E^{2}=\left(j+\frac{1}{2}\right)^{2},
\end{equation}
where $j=\frac{\left(2l-1\right)}{2}$
 with positive integer $l$. In general, the function
 $_{s}Y_{jm}(\theta,\varphi)$ are defined in the following ways:
\begin{eqnarray}
_{s}Y_{jm}(\theta,\varphi)&=&\left(\frac{2j+1}{4\pi}\right)^{\frac{1}{2}}D^{j}_{-sm}(\varphi,\theta,0)\nonumber\\
&=&\left[\frac{(2j+1)}{4\pi}\frac{(j+m)!}{(j+s)!}\frac{(j-m)!}{(j-s)!}\right]^{\frac{1}{2}}
\left(\sin\frac{\theta}{2}\right)^{2j}e^{im\varphi}\nonumber\\
&&\times\sum_{n}\begin{pmatrix}j-s\\n\end{pmatrix}\begin{pmatrix}j+s\\n+s-m\end{pmatrix}(-1)^{j-s-n}
\left(\cot\frac{\theta}{2}\right)^{2n+s-m},
\end{eqnarray}
where $D^{j}_{-sm}(\varphi,\theta,0)$ are the elements of the matrix representations of
the rotation group which satisfy
\begin{eqnarray}
&&D^{j_{1}}_{\mu_{1}m_{1}}(\alpha,\beta,\gamma)D^{j_{2}}_{\mu_{2}m_{2}}(\alpha,\beta,\gamma)
\nonumber\\&&=\sum_{j}\left\langle\;j_{1}j_{2}\mu_{1}\mu_{2}|j\mu_{1}+\mu_{2}\right\rangle\left\langle\;
j_{1}j_{2}m_{1}m_{2}|jm_{1}+m_{2}\right\rangle\;D^{j}_{\mu_{1}+\mu_{2},m_{1}+m_{2}}(\alpha,\beta,\gamma),\\
&&\int\;d\Omega'\;D^{j_{1}\ast}_{\mu_{1}m_{1}}(\alpha,\beta,\gamma)D^{j_{2}}_{\mu_{2}m_{2}}(\alpha,\beta,\gamma)
=\frac{8\pi^{2}}{2j_{1}+1}\delta_{j_{1}j_{2}}\delta_{\mu_{1}\mu_{2}}\delta_{m_{1}m_{2}},
\end{eqnarray}
where $\left\langle\; j_{1}j_{2}m_{1}m_{2}|jm_{1}+m_{2}\right\rangle$ are the
Clebsch-Gordon coefficients and
\begin{eqnarray}
j=j_{1}+j_{2},j_{1}+j_{2}-1,\cdots,|j_{1}-j_{2}|,\\
\int\;d\Omega'\equiv\int^{2\pi}_{0}d\alpha\int^{2\pi}_{0}d\gamma\int^{\pi}_{0}\sin\beta\;d\beta.
\end{eqnarray}
The function $_{s}Y_{jm}(\theta,\varphi)$ satisfies the parity and the orthogonality
relations
\begin{eqnarray}
 _{s}Y^{\ast}_{jm}(\theta,\varphi)=(-1)^{s+m}\;_{-s}Y_{j-m}(\theta,\varphi),\\
\int\;_{s}Y^{\ast}_{j'm'}(\theta,\varphi)\;_{s}Y_{jm}(\theta,\varphi)d\Omega=\delta_{j'm'}\delta_{jm},\\
 \int\;d\Omega=\int^{\pi}_{\theta=0}\int_{\varphi=0}^{2\pi}\sin\theta
 d\theta d\varphi.
\end{eqnarray}
With the customary definition,
\begin{equation}
_{s}Y_{jm}(\theta,\varphi)=\left[\frac{(2j+1)}{4\pi}\frac{(j-m)!}{(j+m)!}\right]^{\frac{1}{2}}\;_{s}P_{jm}
(\theta)e^{im\varphi},
\end{equation}
the operators ${\cal{L}}^{+}_{-s}$ and ${\cal{L}}_{+s}$, for $a\omega=0$, are 'raising'
and 'lowering' operators:
\begin{eqnarray}
\left(\partial_{\theta}-m\csc\theta-s\cot\theta\right)_{s}P_{jm}=-\left[(j-s)(j+s+1)\right]^{\frac{1}{2}}\;_{s+1}P_{jm}\\
\left(\partial_{\theta}+m\csc\theta+s\cot\theta\right)_{s}P_{jm}=+\left[(j+s)(j-s+1)\right]^{\frac{1}{2}}\;_{s-1}P_{jm}.
\end{eqnarray}
\\
\hspace*{7.5mm}For small values of $a$ and $\Lambda$, we can view
$H'={\cal{D}}_{1}+\left(\frac{1}{3}a^{2}\cos^{2}\theta{\cal{D}}_{0}+{\cal{D}}_{2}\right)\Lambda$
 as a perturbation operator and obtain the results by using ordinary perturbation
 theory, say by the first-order expansion
\begin{equation}\label{lambda}
\lambda^{2}=\left(j+\frac{1}{2}\right)^{2}-\left\langle-\frac{1}{2}jm\right|{\cal{D}}_{1}+
\left(\frac{1}{3}a^{2}\cos^{2}\theta{\cal{D}}_{0}+{\cal{D}}_{2}\right)\Lambda\left|-\frac{1}{2}jm\right\rangle+\cdots.
\end{equation}
here
\begin{eqnarray}
&&\left\langle-\frac{1}{2}jm\right|{\cal{D}}_{1}+
\left(\frac{1}{3}a^{2}\cos^{2}\theta{\cal{D}}_{0}+{\cal{D}}_{2}\right)\Lambda\left|-\frac{1}{2}jm\right\rangle\nonumber\\
&&\equiv\int\;_{-\frac{1}{2}}Y_{jm}^{\ast} \left({\cal{D}}_{1}+
\left(\frac{1}{3}a^{2}\cos^{2}\theta{\cal{D}}_{0}+{\cal{D}}_{2}\right)\Lambda\right)\;_{-\frac{1}{2}}Y_{jm}d\Omega\\
&&=\left[2a\omega\;m+\left(\frac{4}{3}a^{3}\omega\;m-\frac{2}{3}a^{2}m^{2}\right)\Lambda\right]
\left\langle-\frac{1}{2}jm\right|\left.-\frac{1}{2}jm\right\rangle\nonumber\\
&&\;\;\;\;-\left[a^{2}\omega^{2}+\left(\frac{2}{3}a^{4}\omega^{2}-\frac{1}{6}a^{2}\right)\Lambda\right]
\left\langle-\frac{1}{2}jm\right|\sin^{2}\theta\left|-\frac{1}{2}jm\right\rangle\nonumber\\
&&\;\;\;\;+\left[\frac{1}{3}a^{2}\left(j+\frac{1}{2}\right)^{2}-\frac{2}{3}a^{3}\omega\;m-\frac{1}{2}a^{2}\right]\Lambda
\left\langle-\frac{1}{2}jm\right|\cos^{2}\theta\left|-\frac{1}{2}jm\right\rangle\nonumber\\
&&\;\;\;\;-\left(a\omega+\frac{1}{3}a^{3}\omega\Lambda\right)
\left\langle-\frac{1}{2}jm\right|\cos\theta\left|-\frac{1}{2}jm\right\rangle\nonumber\\
&&\;\;\;\;-\frac{2}{3}a^{2}\Lambda
\left\langle-\frac{1}{2}jm\right|\cos\theta\sin\theta\;\partial_{\theta}\left|-\frac{1}{2}jm\right\rangle\nonumber\\
&&\;\;\;\;+\frac{1}{3}a^{4}\omega^{2}\Lambda
\left\langle-\frac{1}{2}jm\right|\sin^{2}\theta\cos^{2}\theta\left|-\frac{1}{2}jm\right\rangle\nonumber\\
&&\;\;\;\;-\frac{1}{3}a^{3}\omega\Lambda
\left\langle-\frac{1}{2}jm\right|\cos\theta\sin^{2}\theta\left|-\frac{1}{2}jm\right\rangle.
\end{eqnarray}

\begin{eqnarray}
&&\left\langle-\frac{1}{2}jm\right|\cos\theta\left|-\frac{1}{2}jm\right\rangle=
\left\langle\;j1m0\right|\left.jm\right\rangle\left\langle\;j1\frac{1}{2}0\right|\left.j\frac{1}{2}\right\rangle
=\frac{m}{2}\frac{1}{j(j+1)},\\
&&\left\langle-\frac{1}{2}jm\right|\cos^{2}\theta\left|-\frac{1}{2}jm\right\rangle=\frac{1}{3}+\frac{2}{3}\left\langle
\;j2m0\right|\left.jm\right\rangle\left\langle\;j2\frac{1}{2}0\right|\left.j\frac{1}{2}\right\rangle\nonumber\\
&&=\frac{1}{3}+\frac{2}{3}\frac{\left[3m^{2}-j(j+1)\right]\left[\frac{3}{4}-j(j+1)\right]}{(2j-1)j(j+1)(2j+3)},\\
&&\left\langle-\frac{1}{2}jm\right|\sin^{2}\theta\left|-\frac{1}{2}jm\right\rangle=\frac{2}{3}-\frac{2}{3}\left
\langle\;j2m0\right|\left.jm\right\rangle\left\langle\;j2\frac{1}{2}0\right|\left.j\frac{1}{2}\right\rangle\nonumber\\
&&=\frac{2}{3}-\frac{2}{3}\frac{\left[3m^{2}-j(j+1)\right]\left[\frac{3}{4}-j(j+1)\right]}{(2j-1)j(j+1)(2j+3)},\\
&&\left\langle-\frac{1}{2}jm\right|\cos^{2}\theta\sin^{2}\theta\left|-\frac{1}{2}jm\right\rangle\nonumber\\
&&=\frac{2}{15}+\frac{2}{21}\left\langle\;j2m0\right|\left.jm\right\rangle\left\langle\;j2\frac{1}{2}0\right|
\left.j\frac{1}{2}\right\rangle-\frac{8}{35}\left\langle\;j4m0\right|\left.jm\right\rangle
\left\langle\;j4\frac{1}{2}0\right|\left.j\frac{1}{2}\right\rangle\nonumber\\
&&=\frac{2}{15}+\frac{2}{21}\frac{\left[3m^{2}-j(j+1)\right]\left[\frac{3}{4}-j(j+1)\right]}{(2j-1)j(j+1)(2j+3)}-
\frac{8}{35}\frac{(2j+1)(2j-4)!}{(2j+5)!}\nonumber\\
&&\;\;\;\;\times\left\{(j+m)(j+m-1)[(j+m-2)(6j-8m+3)-9(j-m)(j-3m)]\right.\nonumber\\
&&\;\;\;\;\left.-(j-m)(j-m-1)[9(j+m)(j+3m)-(j-m-2)(6j+8m+3)]\right\}\nonumber\\
&&\;\;\;\;\times\left\{\left(j+\frac{1}{2}\right)\left(j-\frac{1}{2}\right)\left[\left(j-\frac{3}{2}\right)
\left(6j-1\right)-9\left(j-\frac{1}{2}\right)\left(j-\frac{3}{2}\right)\right]\right.\nonumber\\
&&\;\;\;\;\left.-\left(j-\frac{1}{2}\right)\left(j-\frac{3}{2}\right)\left[9\left(j+\frac{1}{2}\right)
\left(j+\frac{3}{2}\right)-\left(j-\frac{5}{2}\right)\left(6j+7\right)\right]\right\},\\
&&\left\langle-\frac{1}{2}jm\right|\cos{\theta}\sin^{2}\theta\left|-\frac{1}{2}jm\right\rangle\nonumber\\
&&=\frac{2}{5}\left\langle\;j1m0\right|\left.jm\right\rangle
\left\langle\;j1\frac{1}{2}0\right|\left.j\frac{1}{2}\right\rangle-\frac{2}{5}\left\langle\;j3m0\right|\left.jm\right
\rangle\left\langle\;j3\frac{1}{2}0\right|\left.j\frac{1}{2}\right\rangle\nonumber\\
&&=\frac{m}{5}\frac{1}{j(j+1)}-\frac{8}{5}\frac{(2j+1)(2j-3)!}{(2j+4)!}\nonumber\\
&&\times\left[\left(j+m\right)\left(j+m-1\right)\left(4j-5m+1\right)
-\left(j-m\right)\left(j-m-1\right)\left(4j+5m+1\right)\right]\nonumber\\
&&\times\left[\left(j+\frac{1}{2}\right)\left(j-\frac{1}{2}\right)\left(4j-\frac{3}{2}\right)
-\left(j-\frac{1}{2}\right)\left(j-\frac{3}{2}\right)\left(4j+\frac{7}{2}\right)\right],\\
&&\left\langle-\frac{1}{2}jm\right|\sin\theta\cos\theta\partial_{\theta}\left|-\frac{1}{2}jm\right\rangle
=\frac{1}{2}\left[\left(j+\frac{1}{2}\right)\sqrt{\frac{2}{3}}\left\langle\;j2m0
\right|\left.jm\right\rangle\left\langle\;j2-\frac{1}{2}1\right|\left.j\frac{1}{2}\right\rangle\right.\nonumber\\
&&\;\;\;\;\left.+\sqrt{\frac{2}{3}\left(j-\frac{1}{2}\right)\left(j+\frac{3}{2}\right)}\left\langle\;j2m0
\right|\left.jm\right\rangle\left\langle\;j2\frac{3}{2}-1\right|\left.j\frac{1}{2}\right\rangle\right]\nonumber\\
&&=\frac{3m^2-j(j+1)}{4j(j+1)}.
\end{eqnarray}
\newpage
\section{QNMS OF KERR-NEWMAN-DE SITTER BLACK HOLE}\setcounter{table}{0}\label{qml}
\begin{table}[htbp]
\begin{center}
\caption{QNMs of massless neutrino for $\Lambda=0.01,Q=0.1,a=0.1$}
\begin{tabular}{c c c c c c c c c c}
 \hline\emph{l}&\emph{n}&\emph{m}$=\frac{1}{2}$&\emph{m}$=-\frac{1}{2}$&\emph{m}
$=\frac{3}{2}$&\emph{m}$=-\frac{3}{2}$&\emph{m}$=\frac{5}{2}$\\
\hline
1&0&0.1851-0.1002\emph{i}&0.1748-0.09909\emph{i}&&&\\
2&0&0.3723-0.09453\emph{i}&0.3639-0.09442\emph{i}&0.3792-0.09474\emph{i}&0.3535-0.09454\emph{i}&\\
&1&0.3729-0.2832\emph{i}&0.3634-0.2837\emph{i}&0.3808-0.2832\emph{i}&0.3521-0.2853\emph{i}&\\
3&0&0.5561-0.09302\emph{i}&0.5481-0.09301\emph{i}&0.5632-0.09310\emph{i}&0.5388-0.09310\emph{i}&0.5694-0.09320\emph{i}\\
&1&0.5564-0.2789\emph{i}&0.5478-0.2792\emph{i}&0.56403-0.2789\emph{i}&0.5379-0.2798\emph{i}&0.5707-0.2790\emph{i}\\
&2&0.5568-0.4646\emph{i}&0.5475-0.4656\emph{i}&0.5651-0.4640\emph{i}&0.5371-0.4673\emph{i}&0.5726-0.4638\emph{i}\\
4&0&0.7399-0.09251\emph{i}&0.7320-0.09251\emph{i}&0.7471-0.09254\emph{i}&0.7232-0.09256\emph{i}&0.7536-0.09260\emph{i}\\
&1&0.7401-0.2775\emph{i}&0.7318-0.2776\emph{i}&0.7476-0.2774\emph{i}&0.7226-0.2779\emph{i}&0.7544-0.2775\emph{i}\\
&2&0.7404-0.4623\emph{i}&0.7316-0.4628\emph{i}&0.7484-0.4620\emph{i}&0.7219-0.4637\emph{i}&0.7557-0.4618\emph{i}\\
&3&0.7407-0.6469\emph{i}&0.7314-0.6482\emph{i}&0.7492-0.6461\emph{i}&0.7213-0.6499\emph{i}&0.7572-0.6455\emph{i}\\
5&0&0.9238-0.09228\emph{i}&0.9159-0.09228\emph{i}&0.9310-0.09230\emph{i}&0.9073-0.09232\emph{i}&0.9377-0.09233\emph{i}\\
&1&0.9239-0.2768\emph{i}&0.9158-0.2769\emph{i}&0.9314-0.2768\emph{i}&0.9069-0.2771\emph{i}&0.9383-0.2769\emph{i}\\
&2&0.9241-0.4612\emph{i}&0.9156-0.4616\emph{i}&0.9320-0.4611\emph{i}&0.9064-0.4621\emph{i}&0.9393-0.4610\emph{i}\\
&3&0.9243-0.6456\emph{i}&0.9154-0.6463\emph{i}&0.9326-0.6451\emph{i}&0.9058-0.6474\emph{i}&0.9404-0.6447\emph{i}\\
&4&0.9246-0.8298\emph{i}&0.9152-0.8312\emph{i}&0.9333-0.8288\emph{i}&0.9053-0.8329\emph{i}&0.9415-0.8281\emph{i}\\
\hline\hline
 \emph{l}&\emph{n}&\emph{m}$=-\frac{5}{2}$&\emph{m}
$=\frac{7}{2}$&\emph{m}$=-\frac{7}{2}$&\emph{m}$=\frac{9}{2}$&\emph{m}$=-\frac{9}{2}$\\
\hline
3&0&0.5279-0.09336\emph{i}&&&\\
&1&0.5266-0.2811\emph{i}&&&\\
&2&0.5256-0.4699\emph{i}&&&\\
4&0&0.7132-0.09269\emph{i}&0.7595-0.09267\emph{i}&0.7020-0.09295\emph{i}&\\
&1&0.7123-0.2785\emph{i}&0.7606-0.2777\emph{i}&0.7008-0.2795\emph{i}&\\
&2&0.7113-0.4651\emph{i}&0.7625-0.4619\emph{i}&0.6995-0.4672\emph{i}&\\
&3&0.7104-0.6523\emph{i}&0.7645-0.6452\emph{i}&0.6987-0.6554\emph{i}&\\
5&0&0.8979-0.09240\emph{i}&0.9439-0.09238\emph{i}&0.8875-0.09254\emph{i}&0.9497-0.09244\emph{i}&0.8761-0.09277\emph{i}\\
&1&0.8973-0.2774\emph{i}&0.9447-0.2770\emph{i}&0.8867-0.2780\emph{i}&0.9507-0.2771\emph{i}&0.8750-0.2788\emph{i}\\
&2&0.8964-0.4629\emph{i}&0.9460-0.4610\emph{i}&0.8855-0.4641\emph{i}&0.9523-0.4611\emph{i}&0.8736-0.4658\emph{i}\\
&3&0.8955-0.6489\emph{i}&0.9476-0.6445\emph{i}&0.8844-0.6509\emph{i}&0.9543-0.6445\emph{i}&0.8725-0.6535\emph{i}\\
&4&0.8948-0.8351\emph{i}&0.9492-0.8275\emph{i}&0.8836-0.8379\emph{i}&0.9564-0.8272\emph{i}&0.8717-0.8414\emph{i}\\
\hline
\end{tabular}
\end{center}
\end{table}

\end{document}